\documentstyle[preprint,prl,aps,epsf,epsfig]{revtex}

\begin{document}
\draft
\tightenlines

\title{The Resistance of Feynman Diagrams and the Percolation Backbone Dimension
}
\author{H. K. Janssen, O. Stenull and K. Oerding
}
\address{
Institut f\"{u}r Theoretische Physik 
III\\Heinrich--Heine--Universit\"{a}t\\Universit\"{a}tsstra{\ss}e 1\\40225 
D\"{u}sseldorf, Germany
}
\date{\today}
\maketitle

\begin{abstract}
We present a new view of Feynman diagrams for the field theory of transport on 
percolation clusters. The diagrams for random resistor networks are interpreted 
as being resistor networks themselves. This simplifies the field theory 
considerably as we demonstrate by calculating the fractal dimension $D_B$ of the 
percolation backbone to three loop order. Using renormalization group methods we 
obtain $D_B = 2 + \epsilon /21 - 172\epsilon^2 /9261 + 2 \epsilon^3 \left( - 
74639 + 22680 \zeta \left( 3 \right) 
 \right)/4084101$, where $\epsilon = 6-d$ with $d$ being the spatial dimension 
and $\zeta \left( 3 \right) = 1.202057..$.
\end{abstract}

\narrowtext

Percolation has gained a vast amount of interest over the last decades (for a 
review see e.g.\cite{bunde_havlin_91,stauffer_aharony_92}). Though it represents 
the simplest model of a disordered system it has many applications e.g.\ 
polymerization, porous and amorphous materials, thin films, spreading of 
epidemics etc. Consider a $d$--dimensional lattice where each bond is randomly 
occupied with probability $p$ or empty with probability $1-p$. Occupied and 
empty bonds may stand for different physical properties. Assume that occupied 
bonds are electrical conductors whereas empty sites are insulators and that 
currents can flow only between nearest neighbors. Suppose a potential difference 
is applied between two sites $x$ and $x^\prime$ located on the same cluster. In 
general not all bonds do carry non--zero current since there may be dangling 
ends. This gives rise to the notion of the backbone. It is defined as the set of 
bonds which are connected to both $x$ and $x^\prime$ by mutually 
non--intersecting paths. Except for Wheatstone bridge type configurations these 
are the bonds that carry non--zero current. The fractal dimension $D_B$ of the 
backbone is defined near the critical concentration $p_c$ by $M_B \sim 
|x-x^\prime|^{D_B}$, where $M_B$ denotes the average number of bonds (the mass) 
of the backbone.

In this paper we evaluate $D_B$ by renormalized field theory. Our approach is 
based on a field theoretic formulation of the randomly diluted nonlinear 
resistor network by Harris\cite{harris_87} which itself was based on work by 
Stephen\cite{stephen_78} and Harris and Lubensky\cite{harris_lubensky_84}. The 
aim of this letter is to present our new interpretation of Feynman diagrams as 
being resistor networks themselves\cite{stenull_janssen_xxx} and to employ this 
interpretation to derive $D_B$ up to third order in $\epsilon =6-d$. 

Consider a nonlinear generalization of the random resistor network as proposed 
by Kenkel and Straley\cite{kenkel_straley_82}. The bonds between sites $i$ and 
$j$ obey a generalized Ohm's Law
\begin{eqnarray}
V_j - V_i = \rho_{i,j} I_{i,j} |I_{i,j}|^{r-1}
\end{eqnarray}
or equivalently 
\begin{eqnarray}
\sigma_{i,j} \left( V_j - V_i \right) | V_j - V_i |^{s-1} = I_{i,j} \ , 
\end{eqnarray}
where $\sigma_{i,j}$ ($\rho_{i,j}$) is the nonlinear conductance (resistance) of 
the bond, $I_{i,j}$ is the current flowing through the bond and $V_i$ is the 
potential at site $i$. The exponents $r$ and $s$ are describing the nonlinearity 
with $r=s^{-1}$. 

The power $P$ dissipated on the backbone between $x$ and $x^\prime$ of this 
nonlinear network reads
\begin{eqnarray}
\label{power}
P = UI = R_r (x ,x^\prime) |I|^{r+1} \ ,
\end{eqnarray}
where $U$ denotes the voltage between the two ports, $I$ the resulting current 
and $R_r (x ,x^\prime)$ the resistance of the backbone. On the other hand we may 
write
\begin{eqnarray}
P = \sum_{i,j} |V_j - V_i | I_{i,j} = \sum_{i,j} \rho_{i,j} |I_{i,j}|^{r+1} \ ,
\end{eqnarray}
where the sum is taken over all bonds on the cluster. The limit $r \to -1$, 
taken from above, provides for a convenient way of summing up all conductors 
carrying non--zero current:
\begin{eqnarray}
R_{-1} (x ,x^\prime) = \sum_{i,j} \rho_{i,j} \ .
\end{eqnarray}
We restrict ourself to the case that all conductors have identical resistance 
$\rho$. Hence $R_{-1}$ is proportional to $M_B$ and $D_B$ is identical to 
$\lim_{r \to -1}\phi_{r}/\nu$, where $\nu$ is the correlation length exponent 
and $\phi_r$ is the resistance exponent defined by $M_r = \langle \chi (x 
,x^\prime) R_r (x ,x^\prime) \rangle_C / \langle \chi (x ,x^\prime) \rangle_C 
\sim |x-x^\prime |^{\phi_r /\nu}$. $\langle ...\rangle_C$ denotes the average 
over all configurations of the diluted lattice and $\chi (x ,x^\prime)$ is an 
indicator function that takes the value one if $x$ and $x^\prime$ are on the 
same cluster and zero otherwise. 

The resistance $R_r (x ,x^\prime)$ can be obtained by solving the circuit 
equations
\begin{eqnarray}
\label{cirquitEquations}
\sum_j \sigma_{i,j} \left( V_i - V_j \right) |V_i - V_j |^{s-1}  = I_i \ ,
\end{eqnarray}
where $I_i = I \left( \delta_{i,x} - \delta_{i,x^\prime} \right)$. The circuit 
equations may be viewed as a consequence of the variation principle
\begin{eqnarray}
\label{variationPrinciple1}
\frac{\partial}{\partial V_i} \left[ \frac{1}{s+1} P \left( \left\{ V \right\} 
\right) + \sum_j I_j V_j \right] = 0 \ ,
\end{eqnarray}
where $\left\{ V \right\}$ denotes the set of voltages belonging to the sites of 
the backbone. Obviously the backbone may contain closed loops as sub--networks. 
Suppose there are currents $\left\{ I^{(l)} \right\}$ circulating independently 
around these closed loops. Then the power is not only a function of $I$ but also 
of the set of loop currents. Conservation of charge holds for every ramification 
of the backbone and this gives rise to another variation principle:
\begin{eqnarray}
\label{variationPrinciple2}
\frac{\partial}{\partial I^{(l)}} P \left( \left\{ I^{(l)} \right\} , I \right) 
= 0 \ .
\end{eqnarray} 
Eq.~(\ref{variationPrinciple2}) may be used to eliminate the loop currents and 
thus provides us with a method to determine the total resistance of the backbone 
via Eq.~(\ref{power}).

A field theory for the nonlinear random resistor network was set up by 
Harris\cite{harris_87} in analogy to the linear 
model\cite{stephen_78,harris_lubensky_84}. In order to overcome difficulties 
associated with $\langle ...\rangle_C$ one employs the replica 
technique\cite{mezard_parizi_virasoro_87}. The network is replicated $D$--fold: 
$V_x \to \vec{V_x} = \left( V_x^{(1)}, \ldots , V_x^{(D)} \right)$. One 
considers the correlation function $G \left( x, x^\prime ;\vec{\lambda} \right) 
= \left\langle \psi_{\vec{\lambda}}(x)\psi_{-\vec{\lambda}}(x^\prime) 
\right\rangle_{\mbox{\scriptsize{rep}}}$ of $\psi_{\vec{\lambda}}(x) = \exp 
\left( i \vec{\lambda} \cdot \vec{V}_x \right)$ where $\vec{\lambda} \cdot 
\vec{V}_x = \sum_{\alpha} \lambda^{(\alpha )} V_x^{(\alpha )}$ and 
$\vec{\lambda} \neq \vec{0}$:
\begin{eqnarray}
G \left( x, x^\prime ;\vec{\lambda} \right) = \left\langle Z^{-D} \int \prod_j 
\prod_{\alpha =1}^D dV_j^\alpha \exp \left( -\frac{1}{s+1} P \left( \left\{ 
\vec{V} \right\} \right) + i \vec{\lambda} \cdot \left( \vec{V}_x  - 
\vec{V}_{x^\prime} \right) \right) \right\rangle_C \ .
\end{eqnarray}
Here $P \left( \left\{ \vec{V} \right\} \right) = \sum_{i,j,\alpha} \sigma_{i,j} 
|V_i^{(\alpha)} - V_j^{(\alpha)}|^{s+1}$ and $Z$ is the usual normalization. In 
contrast to the linear network $P$ is not quadratic and hence the integration is 
not gaussian. As a working hypothesis we assume that a saddle point 
approximation is justified.  For details and conditions to be imposed on 
$\vec{\lambda}$ see\cite{harris_87}. The saddle point equation is nothing else 
than the variation principle stated in Eq.~(\ref{variationPrinciple1}). Thus the 
maximum of the integrand is determined by the solution of the circuit equations 
(\ref{cirquitEquations}) and, up to an unimportant constant,
\begin{eqnarray}
\label{GenFkt2}
G \left( x, x^\prime ;\vec{\lambda} \right) = \left\langle \exp \left(  
\frac{\Lambda_r}{r+1} R_r \left( x,x^\prime \right) \right) \right\rangle_C = 
\left\langle \chi({\rm{\bf x}}, {\rm{\bf x}}^\prime) \right\rangle_C \left( 1 + 
\frac{\Lambda_r}{r+1} M_r ({\rm{\bf x}}, {\rm{\bf x}}^\prime) + \ldots \right) \ 
,
\end{eqnarray}
where $\Lambda_r=\sum_{\alpha =1}^D \left( - {\lambda^{(\alpha )}}^2 
\right)^{(r+1)/2}$. Note that the limit $D\to 0$ has to be taken before $r\to 
-1$ for Eq.~(\ref{GenFkt2}) to be well defined. Contact to the Potts--model can 
be established by switching to voltage variables $\vec{\theta}= \Delta \theta 
\vec{k}$ taking discrete values on a $D$--dimensional torus, i.e.\ $\vec{k}$ is 
chosen to be an $D$--dimensional integer with $-M < k^{(\alpha)} \leq M$ and 
$k^{(\alpha )}=k^{(\alpha )} \mbox{mod} (2M)$. In this discrete picture there 
are $(2M)^D-1$ independent state variables per lattice site and one introduces 
the Potts--spins 
\begin{eqnarray}
\Phi_{\vec{\theta}} \left( x \right) = (2M)^{-D} \sum_{\vec{\lambda} \neq 
\vec{0}} \exp \left( i \vec{\lambda} \cdot \vec{\theta} \right) 
\psi_{\vec{\lambda}} (x) = \delta_{\vec{\theta}, \vec{\theta}_{x}} - (2M)^{-D} 
\end{eqnarray}
subject to the condition $\sum_{\vec{\theta }} \Phi_{\vec{\theta}} \left( x 
\right) = 0$.

The replication procedure induces the effective Hamiltonian
\begin{eqnarray}
H_{\mbox{\scriptsize{rep}}} =  - \ln \left\langle  \exp \left( - \frac{1}{s+1} P 
\right) \right\rangle_C
\end{eqnarray}
which may be expanded in terms of $\psi$:
\begin{eqnarray}
H_{\mbox{\scriptsize{rep}}} =  - \sum_{x, x^\prime} \sum_{\vec{\lambda} \neq 
\vec{0}} K \left(\vec{\lambda} \right) \psi_{\vec{\lambda}}\left( x \right) 
\psi_{-\vec{\lambda}}\left( x^\prime \right) \ .
\end{eqnarray}
Next the kernel is Taylor expanded in the limit of large $\sigma$:
\begin{eqnarray}
K \left( \vec{\lambda} \right) = \tau - w \Lambda_r \ ,
\end{eqnarray}
with $\tau$ and $w\sim \sigma^{-1}$ being expansion coefficients and higher 
order terms are neglected since $H_{\mbox{\scriptsize{rep}}}$ is decaying 
exponentially. By defining the discrete derivative $\partial / \partial 
\theta^{(\alpha )}$ through
\begin{eqnarray}
- \sum_{\vec{\theta}}  \Phi_{\vec{\theta}} \left( x \right) 
\frac{\partial^2}{{\partial \theta^{(\alpha )}}^2} \Phi_{\vec{\theta}} \left( x 
\right) = \sum_{\vec{\lambda} \neq \vec{0}} {\lambda^{(\alpha )}}^2 
\psi_{\vec{\lambda}}(x) \psi_{-\vec{\lambda}}(x)
\end{eqnarray}
one obtains upon Fourier transformation
\begin{eqnarray}
K \left( \Delta_{\vec{\theta}} \right) = \tau - w \left( \Delta_{\vec{\theta}} 
\right)^{(r+1)/2} \ .
\end{eqnarray}
To set up a field theoretic Hamiltonian $\mathcal H \mathnormal$ we proceed with 
the usual coarse graining step and replace the Potts--spins $\Phi_{\vec{\theta}} 
\left( x \right)$ by the order parameter $\varphi \left( {\rm{\bf x}} 
,\vec{\theta} \right)$ defined on a $d$--dimensional spatial continuum. 
Constructing all possible invariants of the symmetries of the model from 
$\sum_{\vec{\theta}} \varphi \left( {\rm{\bf x}} , \vec{\theta} \right)^p$ ($p$ 
denotes some power $>1$) and gradients thereof leads to the following 
Hamiltonian in spirit of the Landau--Ginzburg--Wilson functional (for details 
see\cite{stenull_janssen_xxx}):
\begin{eqnarray}
\mathcal H \mathnormal = \int d^dx \sum_{\vec{\theta}} \bigg\{ \frac{\tau}{2} 
\varphi^2 + \frac{1}{2} \left( \nabla \varphi \right)^2 - \frac{w}{2} \varphi 
\left( \Delta_{\vec{\theta}} \right)^{(r+1)/2}  \varphi + \frac{g}{6}\varphi^3 
\bigg\} \ ,
\end{eqnarray}
where terms of higher order in the fields have been neglected since they turn 
out to be irrelevant in the renormalization group sense. Note that $\mathcal H 
\mathnormal$ reduces to the usual Potts--model Hamiltonian by setting $w=0$.

Now we set up a diagrammatic expansion. Contributing elements are the vertex 
$-g$ and the propagator
\begin{eqnarray}
\label{propagatorDecomp}
\frac{ 1 - \delta_{\vec{\lambda}, \vec{0}}}{{\rm{\bf p}}^2 + \tau - w \Lambda_r} 
= \frac{1}{{\rm{\bf p}}^2 + \tau - w \Lambda_r} - \frac{\delta_{\vec{\lambda}, 
\vec{0}}}{{\rm{\bf p}}^2 + \tau} \ .
\end{eqnarray}
Eq.~(\ref{propagatorDecomp}) shows that the principal propagator decomposes into 
a propagator carrying $\vec{\lambda}$'s (conducting) and one not carrying 
$\vec{\lambda}$'s (insulating). This allows for a schematic decomposition of 
principal diagrams into sums of diagrams consisting of conducting and insulating 
propagators. Here a new interpretation of the Feynman diagrams 
emerges\cite{stenull_janssen_xxx}. They may be viewed as resistor networks 
themselves with conducting propagators corresponding to conductors and 
insulating propagators to open bonds. Schwinger parameters $s_i$ of conducting 
propagators correspond to resistances $\sigma_i^{-1}$ and the replica variables 
$i\vec{\lambda}_i$ to currents. The replica currents are conserved in each 
vertex and we may write $\vec{\lambda}_i = \vec{\lambda}_i \left( \vec{\lambda} 
, \left\{ \vec{\kappa} \right\} \right)$, where $\vec{\lambda}$ is an external 
current and $\left\{ \vec{\kappa} \right\}$ denotes the set of independent loop 
currents. The $\vec{\lambda}$--dependent part of a diagram can be expressed in 
terms of its power $P$:
\begin{eqnarray}
\exp \left( w \sum_{i} s_i {\Lambda_r}_i \right) = \exp \left( w P \left( 
\vec{\lambda} , \left\{ \vec{\kappa} \right\} \right) \right) \ .
\end{eqnarray} 

The new interpretation suggests an alternative way of computing the Feynman 
diagrams. To evaluate sums over independent loop currents
\begin{eqnarray}
\label{toEvaluate}
\sum_{\left\{ \vec{\kappa} \right\}} \exp \left( w P \left( \vec{\lambda} , 
\left\{ \vec{\kappa} \right\} \right) \right) 
\end{eqnarray}
we employ the saddle point method. Note that the saddle point equation is 
nothing else than the variation principle stated in 
Eq.~(\ref{variationPrinciple2}). Thus solving the saddle point equations is 
equivalent to determining the total resistance $R \left( \left\{ s_i \right\} 
\right)$ of a diagram and the saddle point evaluation of (\ref{toEvaluate}) 
yields
\begin{eqnarray}
\exp \left( R_r \left(  \left\{ s_i \right\} \right) w \Lambda_r \right) \ . 
\end{eqnarray}
A completion of squares in the momenta renders the momentum integrations 
straightforward. Thereafter all diagrams are of the form
\begin{eqnarray}
I \left( {\rm{\bf p}}^2 , \vec{\lambda}^2 \right) &=& I_P \left( {\rm{\bf p}}^2 
\right) + I_W \left( {\rm{\bf p}}^2 \right) w \Lambda_r + \ldots
\nonumber \\
&=& \int_0^\infty \prod_i ds_i \left[ 1 + R_r \left(  \left\{ s_i \right\} 
\right) w \Lambda_r + \ldots \right] D \left( {\rm{\bf p}}^2, \left\{ s_i 
\right\} \right) \ ,
\end{eqnarray}
where $D \left( {\rm{\bf p}}^2, \left\{ s_i \right\} \right)$ is a usual 
integrand of the $\phi^3$--theory. The $\phi^3$--theory was investigated to 
three loop order by de Alcantara Bonfim {\it et al}\cite{alcantara_80} and hence 
the remaining task is to calculate the contributions proportional to $w$.  

In order to check if our working hypothesis holds we performed two loop 
calculations for the cases $r \to 0$ and $r \to \infty$ and compared to known 
results. In the limit $r \to 0$ the resistance between two points becomes 
essentially equal to the length of the shortest paths between these points.
We mapped our diagrams onto those studied by Janssen\cite{janssen_85} and 
obtained exactly the same diagrammatic expansion. Consequently, our result for 
the exponent governing the so-called chemical distance $d_{\mbox{{\scriptsize 
min}}} = 2 - \epsilon /6 - \left[ 937/588 + 45/49 \left( \ln 2 -9/10 \ln 3 
\right)\right] \left( \epsilon /6 \right)^2 + \mathcal O \mathnormal \left( 
\epsilon^3  \right)$, $\epsilon =6-d$, is the same as given in 
\cite{janssen_85}. The limit $r \to \infty$ is related to the red (singly 
connected) bonds. Our calculation gives unity for the corresponding exponent in 
accordance with results by Blumenfeld and Aharony\cite{blumenfeld_aharony_85} 
and de Arcangelis {\sl et al}.\cite{arcangelis_85}. We rate these two loop 
results as a strong indication for the validity of the saddle point approach. 

\begin{figure}[h]
\epsfxsize=8.4cm
\centerline{\epsffile{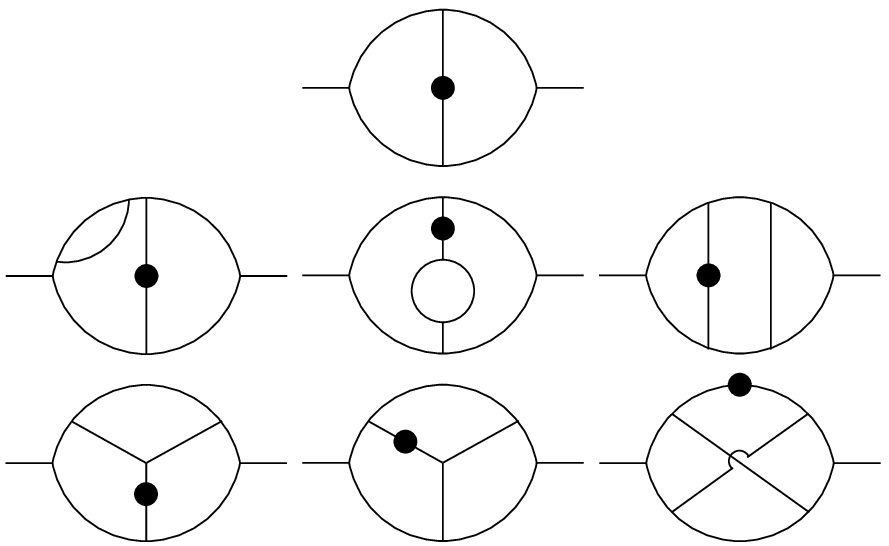}}
\caption[]{\label{theDiagrams}}
\end{figure}

Now we turn to the calculation of $D_B$. In the limit $r \to -1$ only the 
non--planar diagrams listed in Fig.~\ref{theDiagrams} contribute to the 
diagrammatic expansion. We use dimensional regularization and renormalize $w \to 
 Z^{-1} Z_w w$. By employing minimal subtraction to compensate $\epsilon$--poles 
we obtain
\begin{eqnarray}
Z_w = 1 + \frac{u^2}{4\epsilon } + \frac{u^3}{\epsilon^2} \left[ \frac{7}{12} - 
\frac{29}{144}\epsilon - \frac{2}{3} \zeta \left( 3 \right) \epsilon \right] + 
\mathcal O \mathnormal \left( u^4  \right) \ ,
\end{eqnarray}
where $u \propto g^2 \mu^{-\epsilon}$ with $\mu$ being an inverse length scale. 
Note that all non--primitive divergencies are cancelled as renormalizability of 
the perturbation expansion requires. The critical exponents are determined by 
the Wilson--functions $\gamma_{...} \left( u \right) = \mu \frac{\partial 
}{\partial \mu} \ln Z_{...}$ evaluated at the infrared stable fixed point 
$u^\ast$. In particular we are interested in $\eta = \gamma \left( u^\ast 
\right)$ and $\psi = \gamma_w \left( u^\ast\right)$ governing the scaling 
relation
\begin{eqnarray}
\label{scaleRel}
G \left( |{\bf x}-{\bf x}^\prime |; w \right) = l^{d-2+\eta} G \left( l |{\bf 
x}-{\bf x}^\prime|; w/l^{2-\eta +\psi}  \right) \ ,
\end{eqnarray}
where $l$ is a inverse length scale. $\eta$ was calculated to order $\epsilon^3$ 
in\cite{alcantara_80}. For $\psi$ we find
\begin{eqnarray}
\psi = -2 \left( \frac{\epsilon}{7} \right)^2 + \left[ 16 \zeta \left( 3 \right) 
- \frac{2075}{126} \right] \left( \frac{\epsilon}{7} \right)^3 + \mathcal O 
\mathnormal \left( \epsilon^4  \right) \ .
\end{eqnarray}
From Eq.~(\ref{scaleRel}) in conjunction with Eq.~(\ref{GenFkt2}) it follows 
that
\begin{eqnarray}
G \left( |{\bf x}-{\bf x}^\prime |; w \right) = |{\bf x}-{\bf x}^\prime 
|^{2-d-\eta} \left( 1 + w |{\bf x}-{\bf x}^\prime |^{2-\eta +\psi} + \ldots 
\right)
\end{eqnarray}
and hence
\begin{eqnarray}
D_B = 2-\eta +\psi = 2 + \frac{1}{21} \epsilon - \frac{172}{9261} \epsilon^2 + 2 
 \frac{ - 74639 + 22680 \zeta \left( 3 \right) }{4084101} \epsilon^3 + \mathcal 
O \mathnormal \left( \epsilon^4  \right) \ .
\end{eqnarray}
Note that our result agrees to second order in $\epsilon$ with calculations by 
Harris and Lubensky\cite{harris_lubensky_83} based on another approach. This is 
again in favor of our working hypothesis.

\begin{figure}[h]
\epsfxsize=8.4cm
\centerline{\epsffile{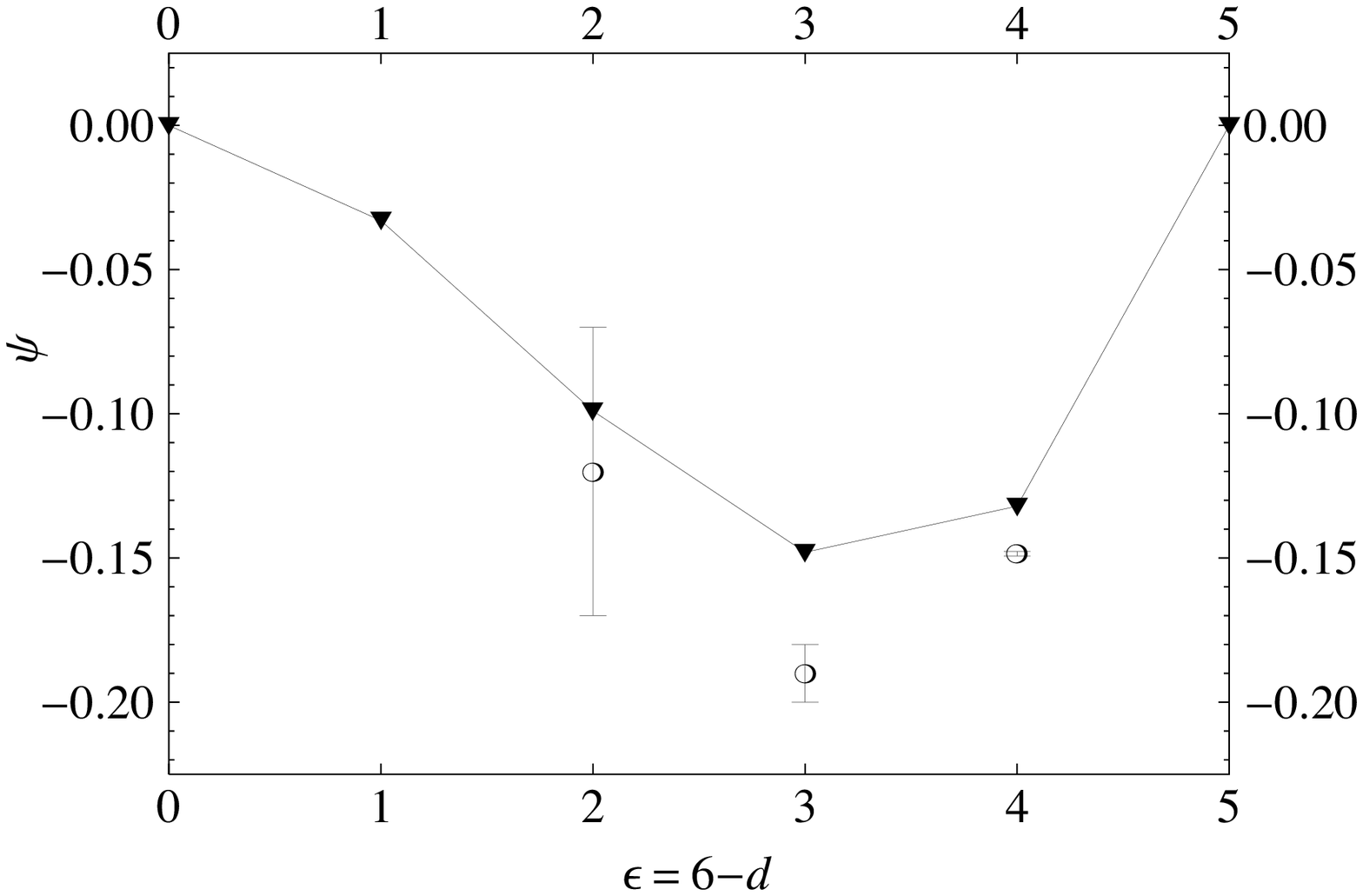}}
\caption[]{\label{data}}
\end{figure}

We compare our result to numerical simulations by 
Grassberger\cite{grassberger_98} and Moukarzel\cite{moukarzel_98}. Due to the 
rich structure of $\eta$ in the percolation problem $\psi$ is better suited for 
such a comparison than $D_B$. It is known exactly that $\psi$ vanishes in one 
dimension. This feature is incorporated by rational approximation yielding
\begin{eqnarray}
\psi \approx - \frac{2 \epsilon^2}{49} \left( 1 - \frac{\epsilon}{5} \right) 
\left( 1 + 1.2625 \frac{\epsilon}{500} \right) \ ,
\end{eqnarray}
which is compared to simulations in Fig.~\ref{data}. For $d=4$ the results agree 
within the numerical errors. However, a higher accuracy of the numerical 
estimate is desirable. For $d=3$ and $d=2$ the analytic result looks less 
realistic and the numerical values are larger. The shape of the dependence of 
$\psi$ on dimensionality is much the same.

We conclude with a few comments. Our interpretation of the Feynman diagrams 
simplifies calculations considerably. The technique used here can be applied
to study other aspects of transport on percolating clusters. In $d=4$ our
result for $D_B$ agrees with recent numerical simulations. For dimensions close 
to 
the upper critical dimension six, our result is the most accurate analytical 
estimate for $D_B$ that we know of.

\acknowledgements
We acknowledge support by the Sonderforschungsbereich 237 ``Unordnung und 
gro{\ss}e Fluktuationen'' of the Deutsche Forschungsgemeinschaft.


\newpage
\begin{figure}[h]
\end{figure}
\noindent
FIG.~\ref{theDiagrams}\newline
The diagrams we computed to determine $D_B$. The lines stand for conducting 
propagators, the solid dots for $\frac{1}{2} \varphi^2$--insertions.
\newline\vspace{0.4cm}\\
FIG.~\ref{data}\newline
Dependence of the exponent $\psi$ on dimensionality. The rational approximation 
(triangles) is compared to numerical results (circles) by Grassberger ($d=2$) 
and Moukarzel ($d=3,4$). They determined $D_B=2-\eta +\psi = \gamma /\nu +\psi$ 
by simulations. For $d=2$ we insert the exact values\cite{nijs_79,nienhuis_82} 
$\nu =4/3$ and $\gamma =43/18$. For $d=3$ we use Monte Carlo results by Ziff and 
Stell\cite{ziff_stell}:  $\nu = 0.875 \pm 0.008$, $\gamma =1.795 \pm 0.005$. For 
$d=4$ we take $\nu^{-1} = 1.44 \pm 0.05$\cite{moukarzel_98} and $\gamma 
=1.44$\cite{stauffer_aharony_92}. 
%
%
\newpage

\end{document}